\documentclass[conference]{IEEEtran}
\usepackage{amsmath}
\usepackage{graphicx}
\usepackage{textcomp}
\usepackage{url}
\usepackage{graphicx}
\usepackage[backend=biber, style=numeric-comp, sorting=none]{biblatex}
\usepackage[hidelinks]{hyperref}
\bibliography{ref}

\begin{document}

\title{Inertial Imaging of Dual Mass Distributions on a Graphene Nanodrum: A Computational Study}

\author{\IEEEauthorblockN{Adhinarayan Naembin Ashok, Sanjam Bedi, Taha Ashraf Ali Shaikh, Jai Aadhithya Ramesh and Adarsh Ganesan}
\IEEEauthorblockA{Department of Electrical and Electronics Engineering,\\Birla Institute of Technology and Science, Pilani - Dubai Campus,\\Dubai International Academic City, Dubai, UAE 345055\\
Email: adarsh@dubai.bits-pilani.ac.in}
}

\maketitle

\begin{abstract}
    This paper presents the possibility for inertial imaging of spatially patterned annular mass distributions of a circular graphene nanodrum resonator. By placing two distinct analytes in concentric annular regions, we harness the vibrational mode-specific sensitivities of the nanodrum to estimate their respective mass densities. An analytical formulation based on the Rayleigh-Ritz principle is developed to relate radial mass loading to modal frequency shifts. Finite element simulations are performed in COMSOL Multiphysics to obtain the shifts in the resonance frequency of vibrational modes under varying geometrical configurations of annular rings. By processing these frequency shifts through a transformation matrix, we estimate the concomitant mass distributions of annular rings. The results indicate that the estimation errors are lower for analytes placed near the antinodal regions of the dominant vibration mode, with the lowest error being 1.82 $\%$ for analyte A and 2.03 $\%$ for analyte B. Furthermore, thinner annular rings demonstrate enhanced detection accuracy due to reduced modal overlap. This study demonstrates an analytical strategy for mass detection using a graphene nanodrum by providing insights into optimal analyte placement and structural design for high-precision multi-target mass sensing applications.\\ 
    
    
\end{abstract}

\section{Introduction}
Mass sensors are employed in a wide range of applications. Using devices based on micro/nanoelectromechanical systems (MEMS/NEMS), one can detect extremely small masses through a readable output of electrical or mechanical signals \cite{ekinci2005nanoelectromechanical}. In chemical and biological detection, functionalized MEMS/NEMS resonators capture specific proteins, DNA strands, or pathogens, and quantify them through adsorption-induced resonance frequency shifts \cite{lavrik2004cantilever}. In thin film deposition monitoring, quartz crystal microbalances (QCMs) provide real-time control over film thickness of materials including metals and polymers, by tracking minute mass changes on the vibrating crystal \cite{lu2012applications}. For environmental monitoring, airborne particulates and trace gas molecules (e.g., volatile organic compounds) are detected, and henceforth enabling continuous pollution assessment \cite{ekinci2005nanoelectromechanical}. In medical diagnostics and drug discovery, NEMS based mass sensors achieved actogram level sensitivity to study biomolecular interactions such as antigen–antibody binding in a label-free manner, thereby facilitating rapid screening of drug candidates \cite{lavrik2004cantilever}\cite{arlett2011comparative}. In another study, ultrasensitive NEMS resonators quantified the mass of single nanoparticles and large biomolecules, pushing the limits of mass resolution towards the zeptogram scale \cite{ekinci2005nanoelectromechanical}.

Two-dimensional (2D) materials of atomic-scale thickness enable ultra-compact nanoelectromechanical sensors. Over the past decade, suspended 2D membrane devices have been proven viable for pressure, acoustic, inertial, mass, and gas sensing \cite{lemme2020nanoelectromechanical}. Dejan Davidovikj et al. \cite{davidovikj2016visualizing} explored the dynamic behavior of suspended graphene nanodrums, whose mechanical properties vary due to static and dynamic wrinkles. Using a phase-sensitive interferometer, they spatially mapped the motion of graphene nanodrum resonators, revealing many interesting spectral features. Kim et al. \cite{kim2018nano} presented an approach for electrical readout of the static displacement of suspended graphene membranes, crucial for developing graphene-based nanomechanical pressure and gas sensors. They used an insulating quartz substrate to enhance the membrane’s contribution to total capacitance and to obtain a reduced gap size of 110 $nm$. With applied pressure, they successfully detected capacitance changes as small as 50 $aF$ and pressure differences as low as 25 mbar. I.E. Rosłoń et al. \cite{roslon2022probing} introduced a novel method using bilayer graphene nanodrums to measure the motion of single bacteria in their aqueous growth environment. Individual E. coli cells were observed to produce random oscillations up to 60 $nm$ in amplitude, exerting forces up to $6 nN$ \cite{roslon2022probing}. The technique also enabled real-time monitoring of changes in nanomotion in response to antibiotics, allowing single-cell antibiotic susceptibility testing. DM Shin et al. \cite{shin2023graphene} introduced a trilayer graphene nanomechanical resonant mass sensor, fabricated via a bottom-up process, achieving sub-actogram resolution at room temperature. These results highlight its promise for high-precision mass measurements in nanobiology and medical diagnostics.

In 2015, Hanay et al. \cite{hanay2015inertial} introduced a nanomechanical sensing technique named 'inertial imaging'. Here, the mass and spatial distribution of individual particles or molecules are analysed through their effect on multiple vibrational modes of a resonator. When a particle adsorbs onto a nanomechanical structure, it changes the device’s natural frequency in a mode-specific manner, depending on both mass and position of the particle. By monitoring the shifts in several resonance modes simultaneously, it becomes possible to infer the total mass as well as the shape and location of the mass being analysed. Unlike optical imaging, inertial imaging is not diffraction-limited, as it relies on the frequency resolution of mechanical modes rather than optical wavelength \cite{chaste2012nanomechanical}. This, for example, makes it exceptionally powerful for analysing nanoparticles, single biomolecules, and viruses, providing information on both morphology and molecular weight \cite{jensen2008atomic}\cite{gil2010nanomechanical}. Ganesan et al. \cite{ganesan2018proposal} showed the possibility to achieve an inertial image with absolute spatial resolution through matrix reconstruction techniques and multiple 2-segment weight matrices.

This paper presents an extension of the inertial imaging approach to simultaneously detect dual mass distributions in a spatially patterned circular graphene nanodrum. In our design, the added mass is distributed in the form of two concentric annular rings on the surface of graphene nanodrum. Each ring is independently functionalized or modified to facilitate binding with different analytes. To study the influence of these mass distributions, we systematically varied both the positions i.e. radial distance from the centre and the thicknesses of mass distribution of the two rings. For each configuration, the eigenfrequencies of the resonator are computed using finite element simulations on COMSOL Multiphysics. Henceforth, the influence of spatial arrangements of mass on the resonance behaviour of multiple vibrational modes is studied. This spatial dependence is then harnessed to estimate the mass distributions. The accuracy of this dual sensing approach is evaluated by comparing the estimated mass values against the actual masses.

The paper is structured as follows: Section II describes the device structure, including the geometric configuration of the graphene nanodrum and the fabrication process to create functionalized concentric annular rings. Section III explains the analytical derivation of the mode-specific frequency shifts using the Rayleigh-Ritz principle and Bessel function approximations. Section IV presents the simulation results, including the eigenfrequency analysis of graphene nanodrum, estimation of errors in the estimation of masses, and the effect of position and thickness of annular rings on sensing precision. Section V summarizes the key findings highlighting the advantages of the proposed configuration, and outlines the future work.

\section{Device Structure}

We consider a suspended circular graphene nanodrum of diameter 16 $\mu m$ and a uniform thickness of 15 $nm$ as shown in Fig. 1. This membrane structure serves as a highly sensitive nanomechanical resonator. Two distinct analytes are introduced onto the surface of the graphene drum, each confined within a separate annular region. Analyte A is localized in a ring-shaped domain defined by the radial bounds $R_1$ and $R_2$, while analyte B occupies a distinct annular region between radii $R_3$ and $R_4$. These nonoverlapping concentric regions are coated with different functional materials. The spatial separation enables the study of individual and localized mass-loading effects on the vibrational behavior of the graphene nanodrum, including frequency shifts and perturbations in mode shapes. This configuration supports multi-analyte detection by harnessing the spatially dependent sensitivity of different vibrational modes.

The fabrication process begins with the wet transfer of a commercially available monolayer graphene (black) onto a Si/SiO\textsubscript{2} substrate (blue). We further coat a Polymethyl Methacrylate (PMMA) polymer resist (yellow) \cite{wei2021molecular}. Using electron beam lithography (EBL), two annular rings are patterned one at a time into the PMMA mask, exposing selected annular graphene regions underneath. This is demonstrated in Fig. 2. In the first functionalization cycle, the exposed annular graphene regions are activated using sodium-potassium (Na/K) ions, making the graphene negatively charged and highly reactive to electrophilic species \cite{wei2021molecular}. Subsequent treatment with benzonitrile (PhCN) in the presence of deuteroxide (D\textsubscript{2}O), results in covalent attachment of the functional group, forming the first functionalized annular zone (purple ring). Now, before proceeding to the next EBL step, PMMA is deposited to protect the first annular zone from the second functionalization step. EBL now creates a concentric ring, providing a new annular exposure zone around the previously functionalized area. Na/K activation followed by exposure to iodine monochloride (ICl) in presence of PhCN enables annular functionalization (green ring), yielding two distinct concentric annular zones, each with different chemical functionalities \cite{wei2021molecular}. Here, we considered covalently bound deutero- and chloro atoms as representative functional materials to demonstrate the fabrication process. However, various other substances can also be considered for functionalization based on the target application.

\begin{figure}
    \centering
    \includegraphics[width=1\linewidth]{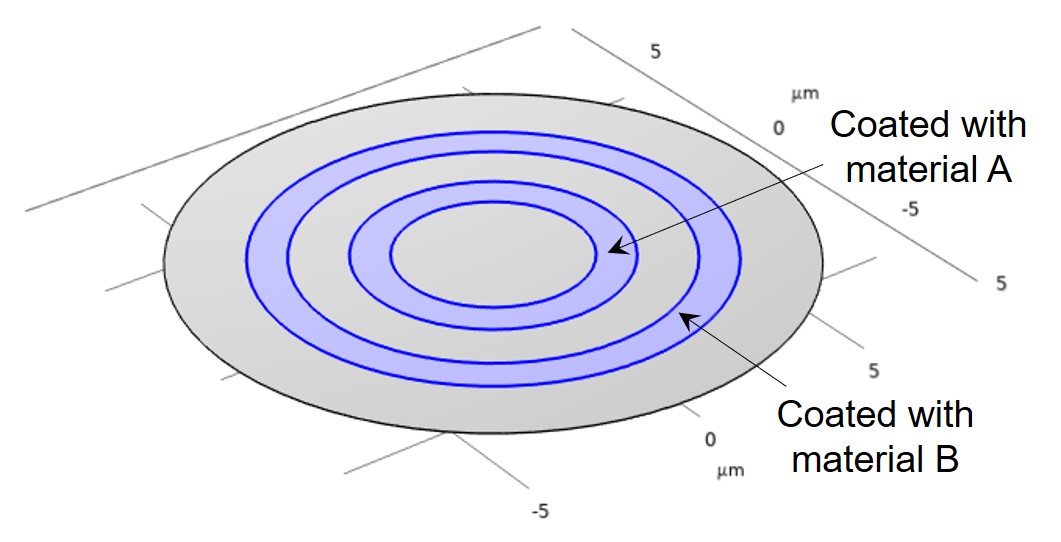}
    \caption{Geometric configuration of a circular graphene nanodrum with functionalized annual rings for detecting analytes A and B}
    \label{fig:enter-label}
\end{figure} 
\begin{figure*}
    \centering
    \includegraphics[width=0.95\linewidth]{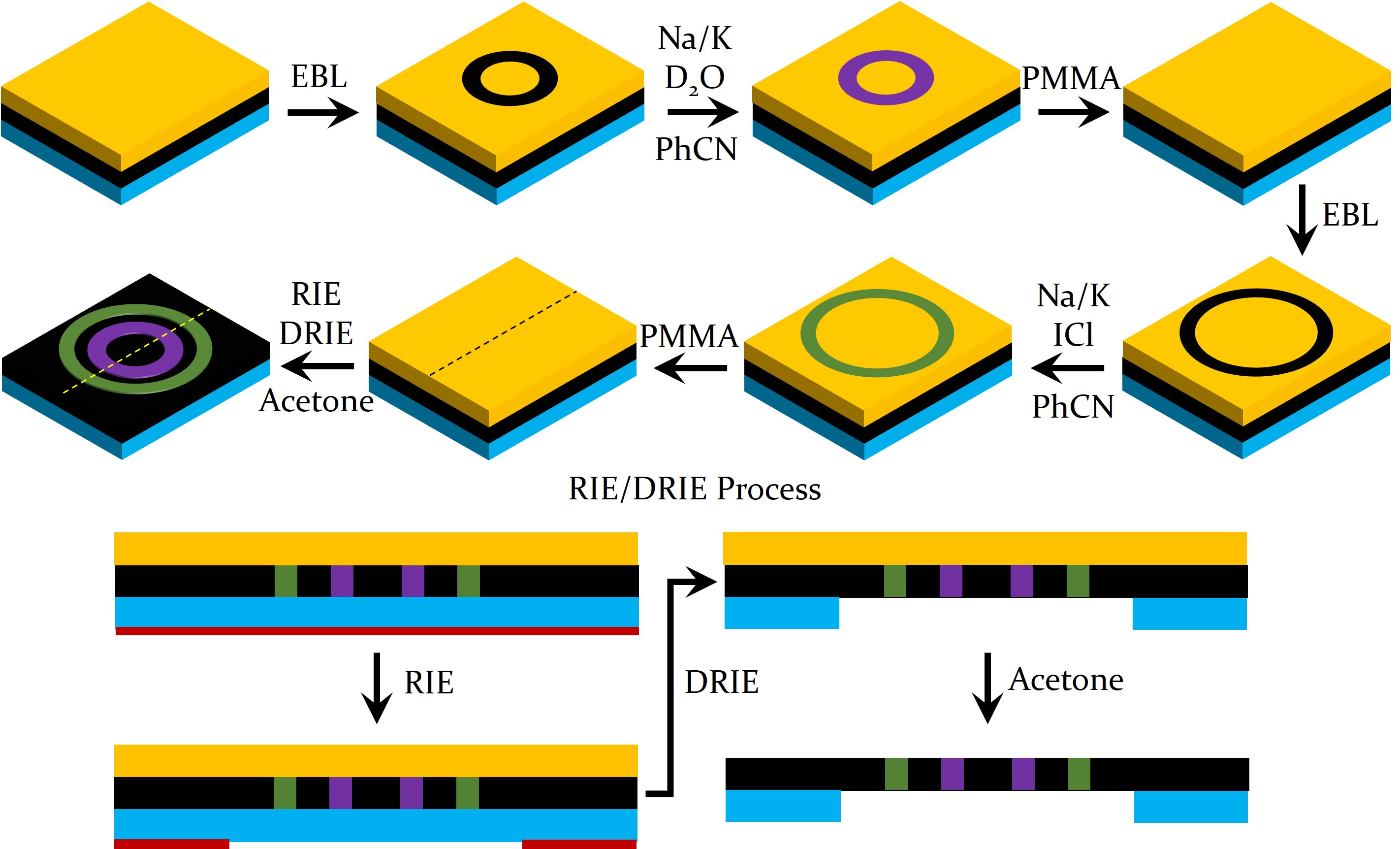}
    \caption{Step by step fabrication procedure for patterned functionalization of graphene. Here, yellow, black and blue regions represent PMMA, graphene and silicon respectively.}
    \label{fig:enter-label}
\end{figure*} 

Now, in order to make a suspended graphene membrane, the substrate layer needs to be etched. Based on the device geometry, selected regions of the protective SiO\textsubscript{2} are patterned using reactive ion etching (RIE), thereby exposing specific areas of the Si substrate. The masking layer protects the underlying substrate regions from further etching to ensure structural support of the device. The exposed Si regions are subsequently etched using a deep reactive ion etching (DRIE) process to achieve the desired vertical profiles. Finally, the PMMA layer is removed by exposure to acetone vapors, which completes the fabrication process \cite{wei2021molecular}.

\section{Analytical Derivation}

Graphene nanodrums possess multiple modes of vibrations. In this work, we considered modes 01 and 02 as shown in Fig. 3. The shapes of these modes can be written as
\begin{equation}
\Phi_n(r) = A_{0n} \, J_0\left( \frac{\alpha_{0n} r}{R} \right)
\end{equation}

Here, $J_0$ is the Bessel function of the first kind of order zero. The mass distribution $\mu(r)$ on nanodrums can cause shifts in the resonance frequencies of these modes, which can be modelled using Rayleigh-Ritz principle.
\begin{equation}
F_n = \frac{\Delta \omega_n}{\omega_n} = \sigma_n \, \frac{\iint \mu(r)\, |\Phi_n(r)|^2 \, dA}{\iint |\Phi_n(r)|^2 \, dA}
\end{equation}

Substituting (1) in (2), we get
\begin{equation}
\begin{aligned}
F_n &= \sigma_n \, \frac{\int_0^R \int_0^{2\pi} \mu(r) \left( A_{0n} J_0\left( \frac{\alpha_{0n} r}{R} \right) \right)^2 r \, d\theta \, dr}
{\int_0^R \int_0^{2\pi} \left( A_{0n} J_0\left( \frac{\alpha_{0n} r}{R} \right) \right)^2 r \, d\theta \, dr} \\
&= \sigma_n \, \frac{\int_0^R \mu(r) J_0^2\left( \frac{\alpha_{0n} r}{R} \right) r \, dr}
{\int_0^R J_0^2\left( \frac{\alpha_{0n} r}{R} \right) r \, dr}
\end{aligned}
\end{equation}

Since the addition of masses only occurs on the functionalized annular rings of nanodrums, we have
\[
\mu(r) = 
\begin{cases}
\mu_1 & \text{for } R_1 \leq r \leq R_2 \\
\mu_2 & \text{for } R_3 \leq r \leq R_4 \\
0     & \text{elsewhere}
\end{cases}
\]

\begin{figure}
    \centering
    \includegraphics[width=0.8\linewidth]{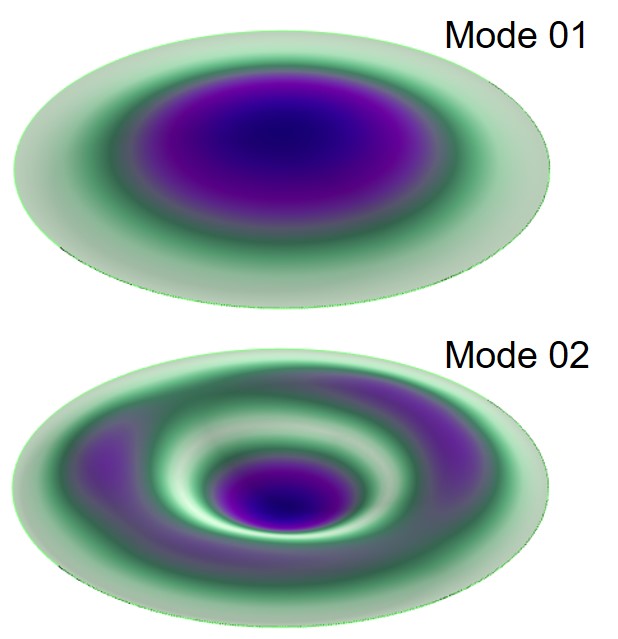}
    \caption{Displacement profiles of modes 01 and 02 of graphene nanodrum}
    \label{fig:enter-label}
\end{figure} 

By substituting $\mu(r)$ in (3), we get
\begin{equation*}
F_n = \sigma_n \, \frac{
\mu_1 \int_{R_1}^{R_2} J_0^2\left( \frac{\alpha_{0n} r}{R} \right) r \, dr 
+ \mu_2 \int_{R_3}^{R_4} J_0^2\left( \frac{\alpha_{0n} r}{R} \right) r \, dr
}{
\int_0^R J_0^2\left( \frac{\alpha_{0n} r}{R} \right) r \, dr
}
\end{equation*}
\begin{equation*}
\begin{aligned}
\Rightarrow \sigma_n \, \Bigg[ &\frac{
\mu_1 \left( \int_0^{R_2} J_0^2\left( \frac{\alpha_{0n} r}{R} \right) r \, dr 
- \int_0^{R_1} J_0^2\left( \frac{\alpha_{0n} r}{R} \right) r \, dr \right)}{
\int_0^R J_0^2\left( \frac{\alpha_{0n} r}{R} \right) r \, dr} \\
&+ \frac{
\mu_2 \left( \int_0^{R_4} J_0^2\left( \frac{\alpha_{0n} r}{R} \right) r \, dr 
- \int_0^{R_3} J_0^2\left( \frac{\alpha_{0n} r}{R} \right) r \, dr \right)}{
\int_0^R J_0^2\left( \frac{\alpha_{0n} r}{R} \right) r \, dr} \Bigg]
\end{aligned}
\end{equation*}
\begin{equation}
\begin{aligned}
\Rightarrow \sigma_n \, \frac{
\mu_1 \left[ K_n(R_2) - K_n(R_1) \right] + 
\mu_2 \left[ K_n(R_4) - K_n(R_3) \right]
}{
K_n(R)
}
\end{aligned}
\end{equation}
\begin{figure}
    \centering
    \includegraphics[width=0.95\linewidth]{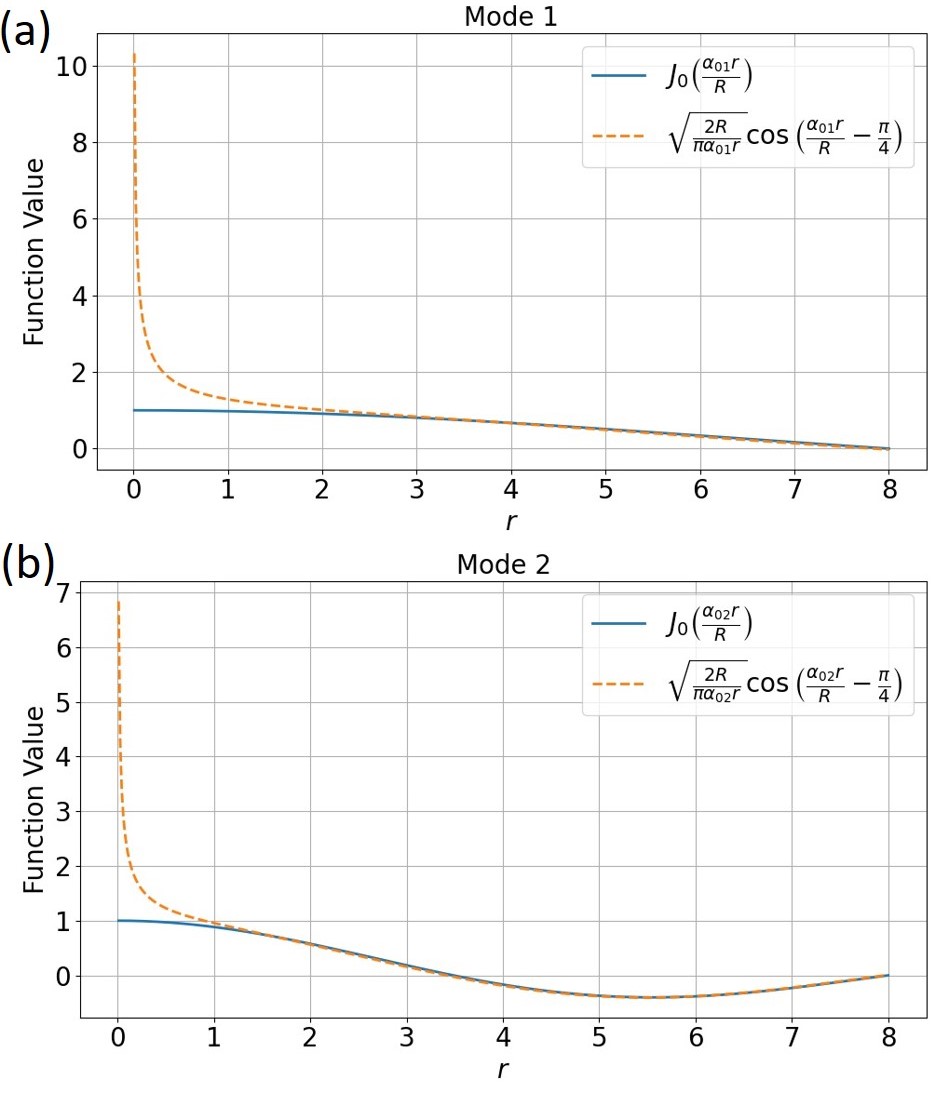}
    \caption{Comparison of the functional values of $J_0$ based on numpy library of python and the corresponding cosine appoximation for: (a) mode 01 (n = 1) (b) mode 02 (n = 2). Here, $r$ is in $\mu m$.}
    \label{fig:enter-label}
\end{figure}
Here, $K_n(R_x) = \int_0^{R_x} r \left[ J_0\left( \frac{\alpha_{0n} r}{R} \right) \right]^2 dr$. Now to find an expression for $K_n(R_x)$, we use a cosine approximation for $J_0$. 
\begin{equation}
J_0\left( \frac{\alpha_{0n} r}{R} \right) = \sqrt{ \frac{2R}{\pi \alpha_{0n} r} } \cos\left( \frac{\alpha_{0n} r}{R} - \frac{\pi}{4} \right)
\end{equation}
Here, $\alpha_{0n} \approx \left(n - \frac{1}{4}\right) \pi$. Fig. 4 shows the validity of the approximation. While the error is high when $r \to 0$ , the approximation is very accurate for higher values of $r$. Now, we find the square of the Bessel function as, 
\begin{equation}
\left[ J_0\left( \frac{\alpha_{0n} r}{R} \right) \right]^2 = \frac{R}{\pi \alpha_{0n} r} + \frac{R}{\pi \alpha_{0n} r} \cos\left( \frac{2\alpha_{0n} r}{R} - \frac{\pi}{2} \right)
\end{equation}
Hence, we have
\begin{equation}
\begin{aligned}
K_n(R_x) &= \int_0^{R_x} r \left[ J_0\left( \frac{\alpha_{0n} r}{R} \right) \right]^2 dr \\
&\approx \int_0^{R_x} \left( \frac{R}{\pi \alpha_{0n}} + \frac{R}{\pi \alpha_{0n}} \cos\left( \frac{2 \alpha_{0n} r}{R} - \frac{\pi}{2} \right) \right) dr \\
&= \frac{R}{\pi \alpha_{0n}} \left[ R_x + \frac{R}{2\alpha_{0n}} \sin\left( \frac{2 \alpha_{0n} R_x}{R} - \frac{\pi}{2} \right) + \frac{R}{2\alpha_{0n}} \right] \\
&= \frac{R}{\pi^2 (n - \tfrac{1}{4})} \left[ R_x + \right. \\
&\quad \frac{R}{2\pi (n - \tfrac{1}{4})} \sin\left( 2\pi (n - \tfrac{1}{4}) \frac{R_x}{R} - \frac{\pi}{2} \right) \\
&\quad \left. + \frac{R}{2\pi (n - \tfrac{1}{4})} \right]
\end{aligned}
\end{equation}

Therefore, by substituting (7) in (4), we get, 
\begin{equation*}
\begin{aligned}
\tilde{F}_n &= \frac{F_n K_n(R)}{\sigma_n} \\
&= \mu_1 \Bigg( R_2 - R_1 
+ \frac{R}{2\pi(n - \tfrac{1}{4})} \Big[ 
\sin\bigg((n - \tfrac{1}{4}) \frac{2\pi R_2}{R} - \frac{\pi}{2} \bigg) \\
&\quad \qquad \qquad \qquad \qquad 
- \sin\bigg((n - \tfrac{1}{4}) \frac{2\pi R_1}{R} - \frac{\pi}{2} \bigg) 
\Big] \Bigg) \\
&\quad + \mu_2 \Bigg( R_4 - R_3 
+ \frac{R}{2\pi(n - \tfrac{1}{4})} \Big[ 
\sin\bigg((n - \tfrac{1}{4}) \frac{2\pi R_4}{R} - \frac{\pi}{2} \bigg) \\
&\quad \qquad \qquad \qquad \qquad 
- \sin\bigg((n - \tfrac{1}{4}) \frac{2\pi R_3}{R} - \frac{\pi}{2} \bigg) 
\Big] \Bigg)
\end{aligned}
\end{equation*}

\begin{equation}
\begin{aligned}
\Rightarrow \mu_1 \Bigg( R_2 - R_1 - \frac{R}{\pi(n - \tfrac{1}{4})} \Bigg[ 
    \cos\left( (n - \tfrac{1}{4}) \cdot \frac{2\pi(R_1 + R_2)}{R} \right)\cdot \\
    \quad \sin\left( (n - \tfrac{1}{4}) \cdot \frac{2\pi(R_2 - R_1)}{R} \right) \Bigg] \Bigg) 
\\
\quad + \mu_2 \Bigg( R_4 - R_3  - \frac{R}{\pi(n - \tfrac{1}{4})} \Bigg[
    \cos\left( (n - \tfrac{1}{4}) \cdot \frac{2\pi(R_3 + R_4)}{R} \right) \\
\quad \cdot \sin\left( (n - \tfrac{1}{4}) \cdot \frac{2\pi(R_4 - R_3)}{R} \right) 
\Bigg] \Bigg)
\end{aligned}
\end{equation}

\begin{figure}
    \centering
    \includegraphics[width=0.95\linewidth]{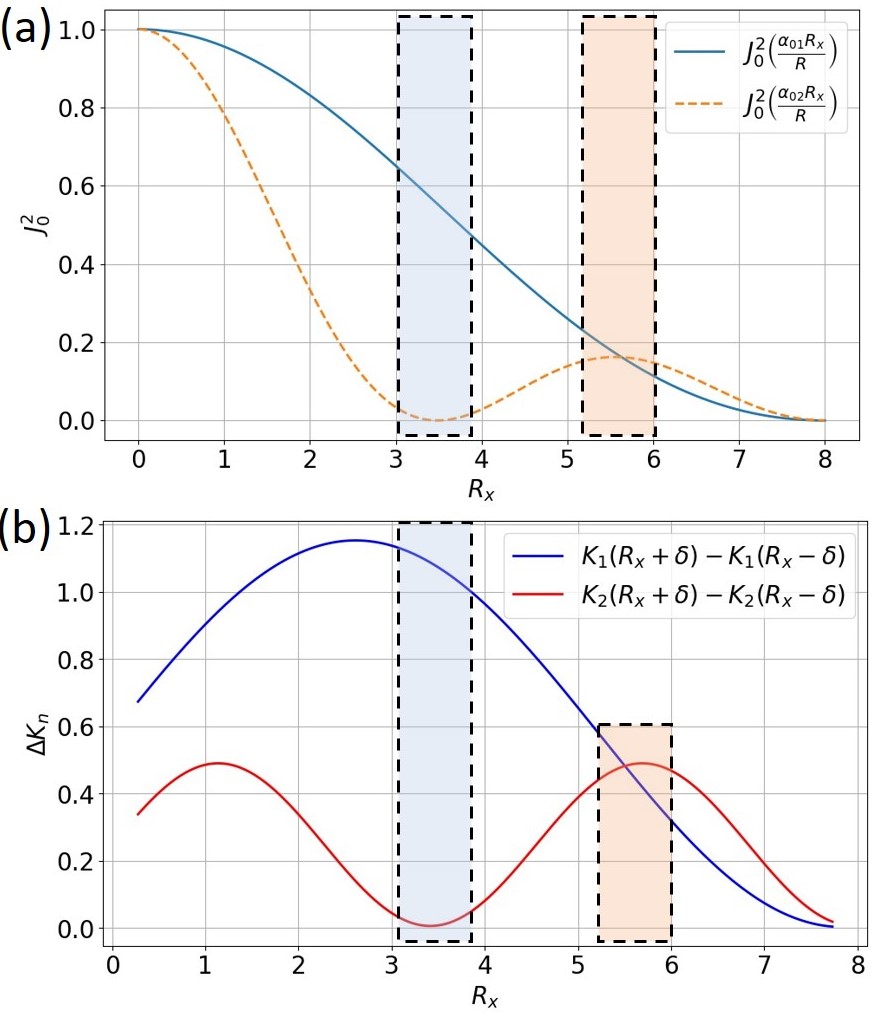}
    \caption{(a-b) Comparison of $J_0$ and $\Delta K_n$ for modes 01 and 02 respectively. Here, $R_x$ is in $\mu m$.}
    \label{fig:enter-label}
\end{figure}

Or simply, we can write \\ 
\begin{equation}
\tilde{F}_1 = \mu_1 \left(K_1(R_2) - K_1(R_1)\right) + \mu_2 \left(K_1(R_4) - K_1(R_3)\right)
\end{equation} 
\begin{equation}
\tilde{F}_2 = \mu_2 \left(K_2(R_2) - K_2(R_1)\right) + \mu_2 \left(K_2(R_4) - K_2(R_3)\right)
\end{equation}
This can be casted in matrix form as \\
\begin{equation*}
\begin{bmatrix}
\tilde{F}_1 \\[6pt]
\tilde{F}_2
\end{bmatrix}
=
\begin{bmatrix}
K_1(R_2) - K_1(R_1) & K_1(R_4) - K_1(R_3) \\[6pt]
K_2(R_2) - K_2(R_1) & K_2(R_4) - K_2(R_3)
\end{bmatrix}
\begin{bmatrix}
\mu_1 \\[6pt]
\mu_2
\end{bmatrix}
\end{equation*}
\begin{equation}
\begin{bmatrix}
\tilde{F}_1 \\[6pt]
\tilde{F}_2
\end{bmatrix} = K \begin{bmatrix}
\mu_1 \\[6pt]
\mu_2
\end{bmatrix}
\end{equation}

\begin{equation}
\begin{bmatrix}
\mu_{1,est} \\[6pt]
\mu_{2,est}
\end{bmatrix}
=
K^{-1}
\begin{bmatrix}
\tilde{F}_1 \\[6pt]
\tilde{F}_2
\end{bmatrix}
\end{equation}

Now, if the frequency shifts are known through experiments, then the unknown mass distribution $\mu_1$ and $\mu_2$ can be estimated by simply inverting the K-matrix. From Fig. 5, it can be observed that for values of $R_x$ between $3 \mu m$ and $4 \mu m$, a node appears in mode 02. As a result, any mass placed will minimally affect its resonance frequency in comparison to that of mode 01. Similarly, for $R_x$ between $5 \mu m$ and $6 \mu m$, a node appears in mode 01, and hence will render mode 01 minimally sensitive to mass. This specific spatial arrangement improves overall mass detection accuracy by minimizing modal interference and ensuring that each analyte aligns with a highly responsive vibrational mode.

\section{Results}

\begin{figure}
    \centering
    \includegraphics[width=0.85\linewidth]{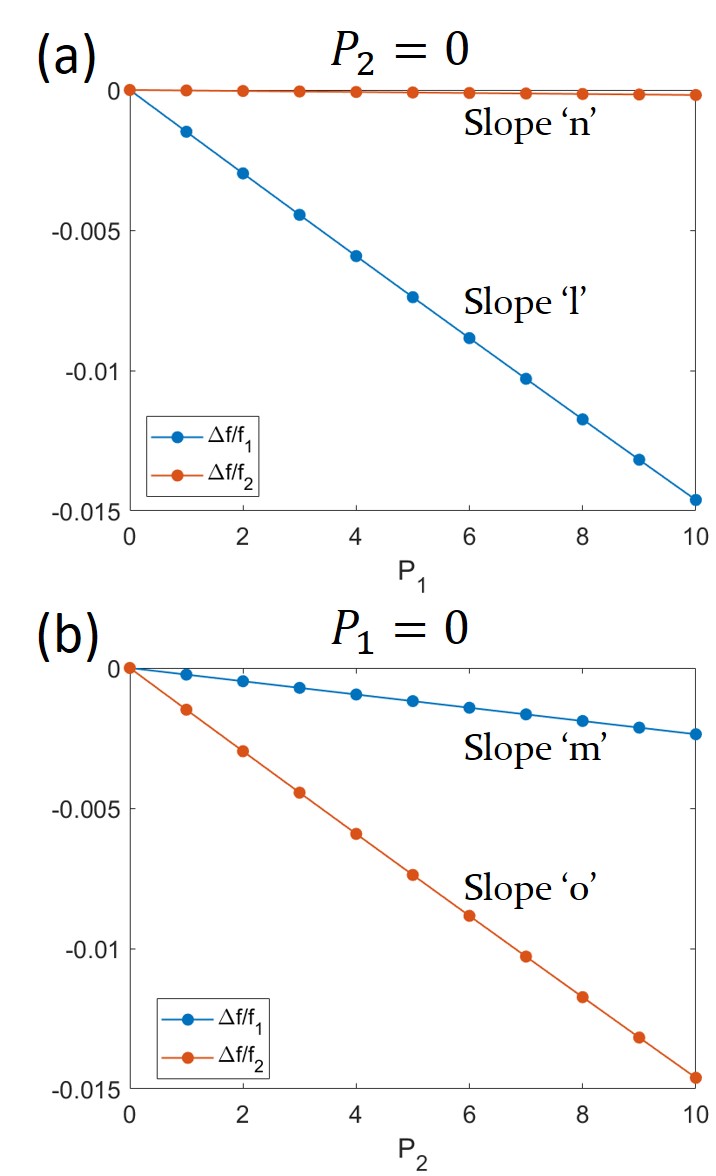}
    \caption{Variation of $\frac{\Delta f_1}{f_1}$  and $\frac{\Delta f_2}{f_2}$ when (a) $P_2 = 0$ and $P_1$ is varied from 0 $kg m^{-2}$ to 10 $kg m^{-2}$ (b) $P_1 = 0$ and $P_2$ is varied from 0 $kg m^{-2}$ to 10 $kg m^{-2}$. Here, $R_1=2.9 \mu m$, $R_2=3.3 \mu m$, $R_3=5.6 \mu m$, $R_4=6 \mu m$ and $R=8 \mu m$.}
    \label{fig:enter-label}
\end{figure}

\begin{figure}
    \centering
    \includegraphics[width=0.90\linewidth]{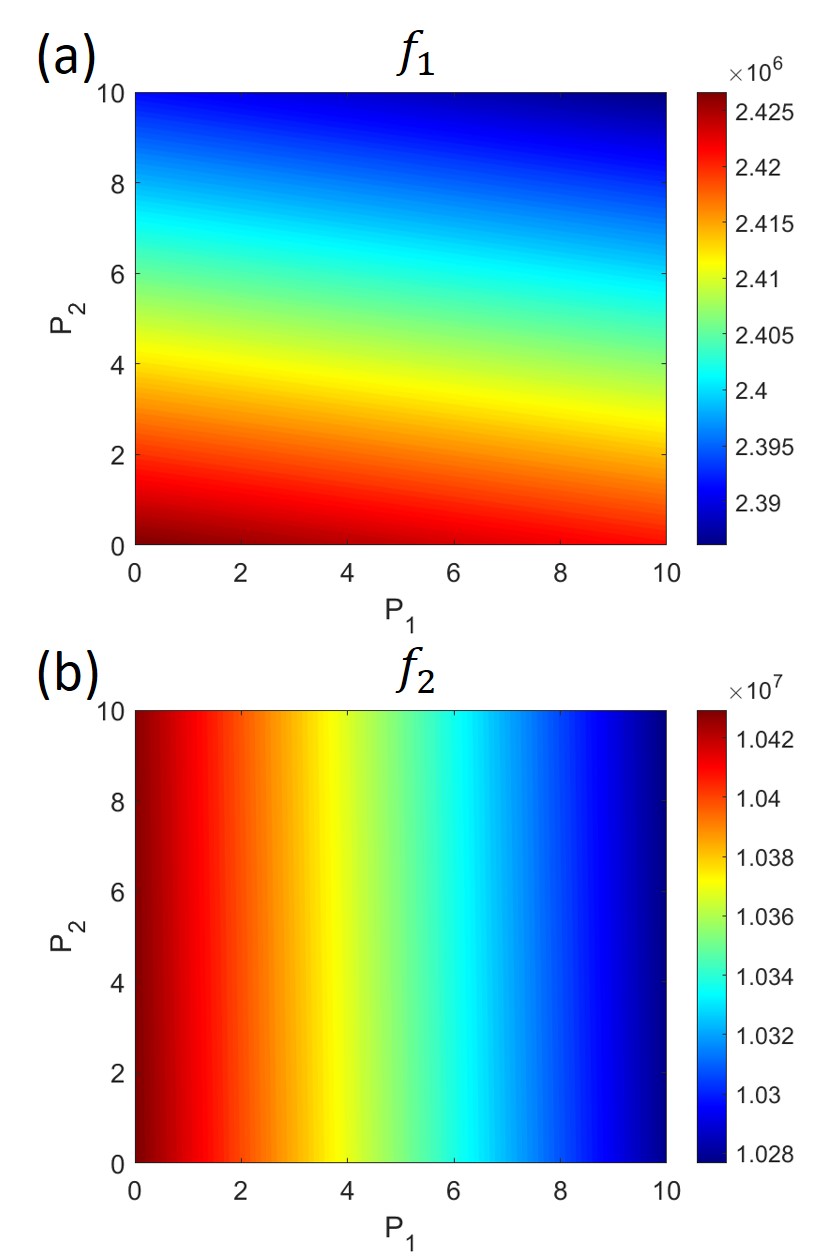}
    \caption{Resonance frequencies (a) $f_1$ and (b) $f_2$, when both $P_1$ and $P_2$ are varied from 0 $kg m^{-2}$ to 10 $kg m^{-2}$. Here, $R_1=2.9 \mu m$, $R_2=3.3 \mu m$, $R_3=5.6 \mu m$, $R_4=6 \mu m$ and $R=8 \mu m$.}
    \label{fig:enter-label}
\end{figure}

\begin{figure}
    \centering
    \includegraphics[width=0.85\linewidth]{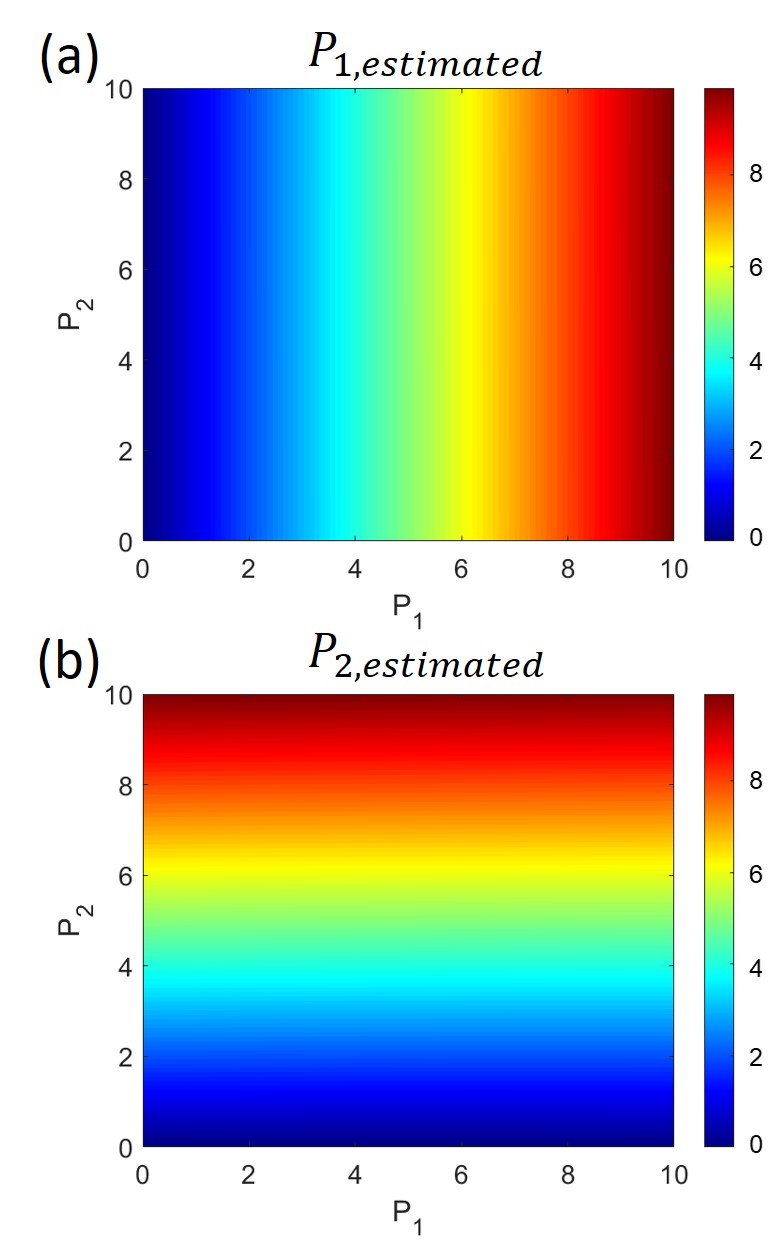}
    \caption{Estimation of (a) $P_1$ and (b) $P_2$ in $kg m^{-2}$ for $P_1$ and $P_2$ in the range 0 $kg m^{-2}$ to 10 $kg m^{-2}$. Here, $R_1=2.9 \mu m$, $R_2=3.3 \mu m$, $R_3=5.6 \mu m$, $R_4=6 \mu m$ and $R=8 \mu m$.}
    \label{fig:enter-label}
\end{figure}

\begin{figure}
    \centering
    \includegraphics[width=0.90\linewidth]{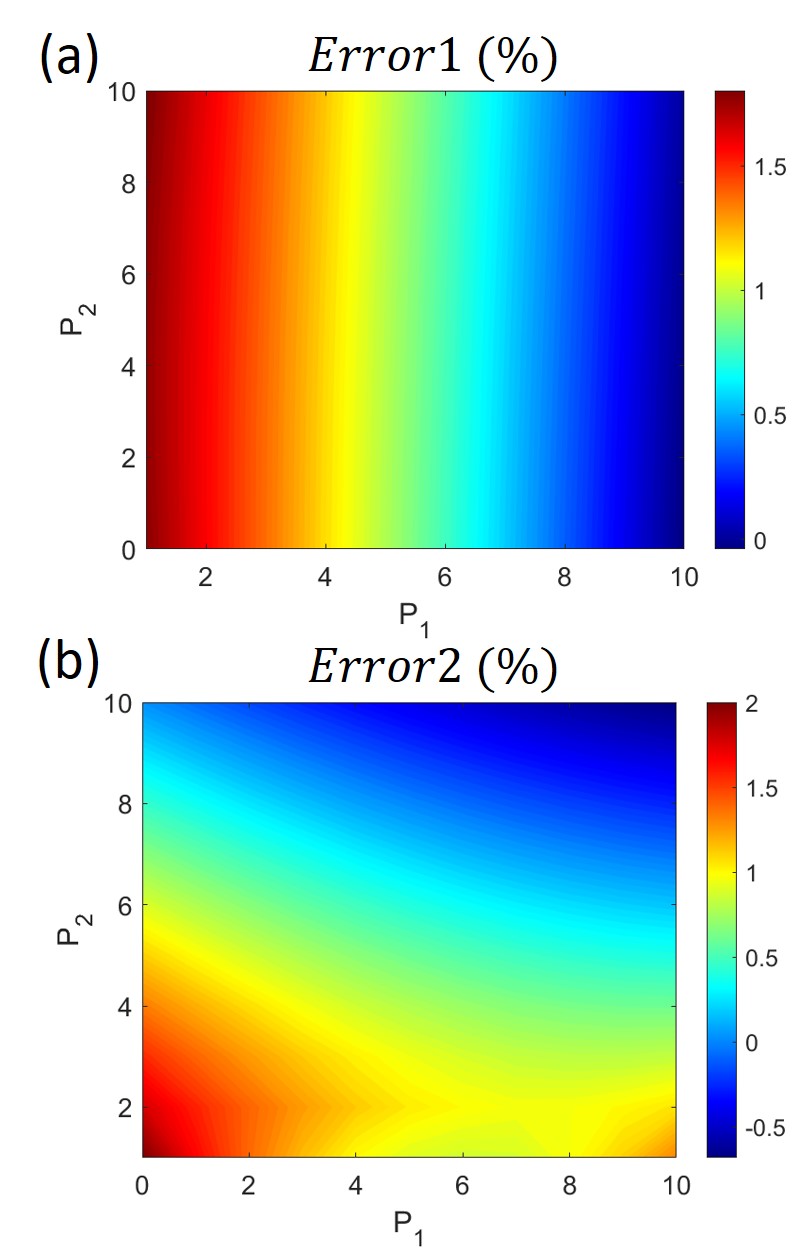}
    \caption{Estimation errors (a) $\frac{P_1-P_{1,est}}{P_1}\%$ and (b) $\frac{P_2-P_{2,est}}{P_2}\%$ for $P_1$ and $P_2$ in the range 0 $kg m^{-2}$ to 10 $kg m^{-2}$. Here, $R_1=2.9 \mu m$, $R_2=3.3 \mu m$, $R_3=5.6 \mu m$, $R_4=6 \mu m$ and $R=8 \mu m$.}
    \label{fig:enter-label}
\end{figure}

\begin{figure}
    \centering
    \includegraphics[width=0.90\linewidth]{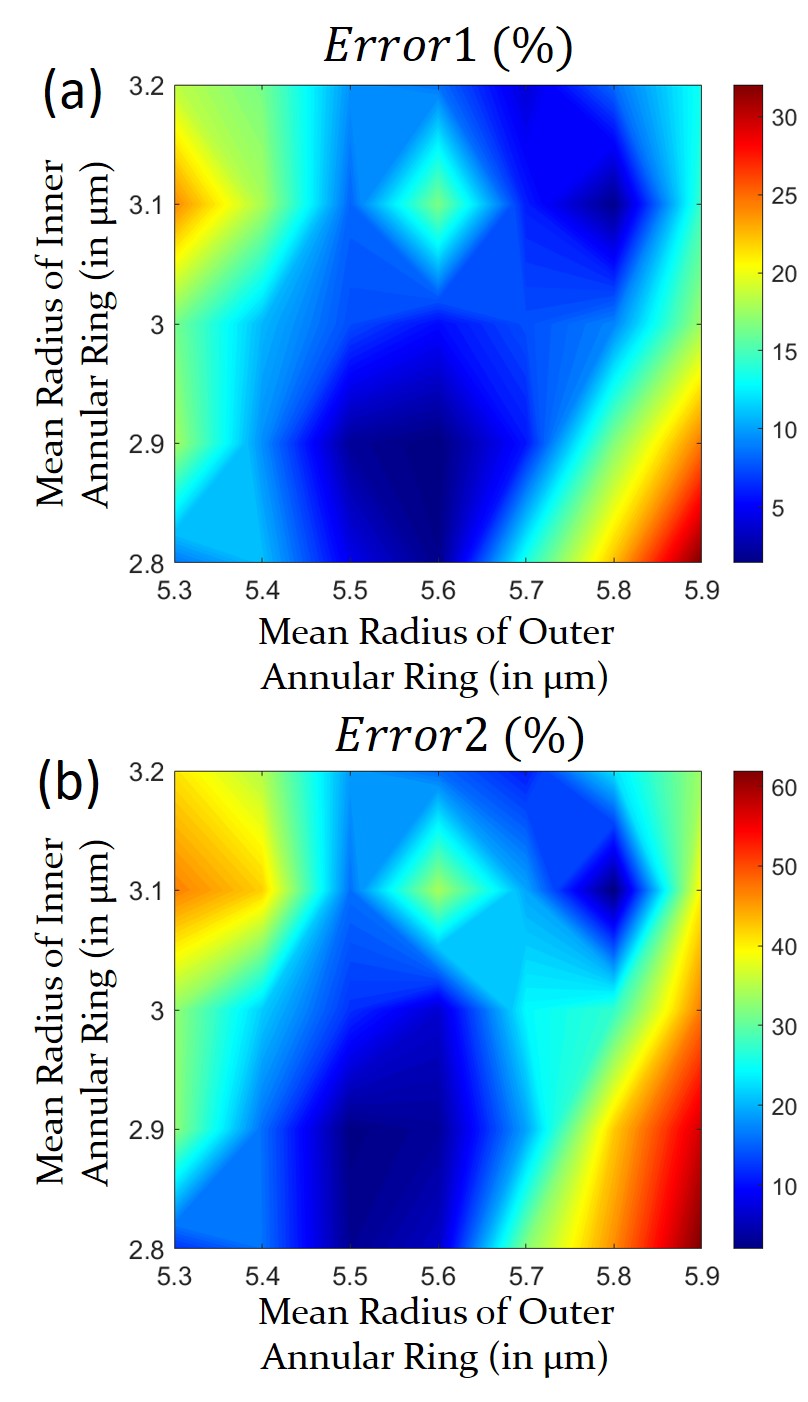}
    \caption{Minimum estimation errors (a) $(\frac{P_1-P_{1,est}}{P_1})_{min}\%$ and (b) $(\frac{P_2-P_{2,est}}{P_2})_{min}\%$ for $P_1$ and $P_2$ in the range 0 $kg m^{-2}$ to 10 $kg m^{-2}$ for different mean radii of inner ring $\frac{R_1+R_2}{2}$ and outer ring $\frac{R_3+R_4}{2}$. Here, $R_1=2.6-3 \mu m$, $R_2=3-3.4 \mu m$, $R_3=5.1-5.7 \mu m$, $R_4=5.5-6.1 \mu m$, $R=8 \mu m$ and $R_2-R_1=R_4-R_3=0.4 \mu m$.}
\end{figure}

\begin{figure}
    \centering
    \includegraphics[width=0.90\linewidth]{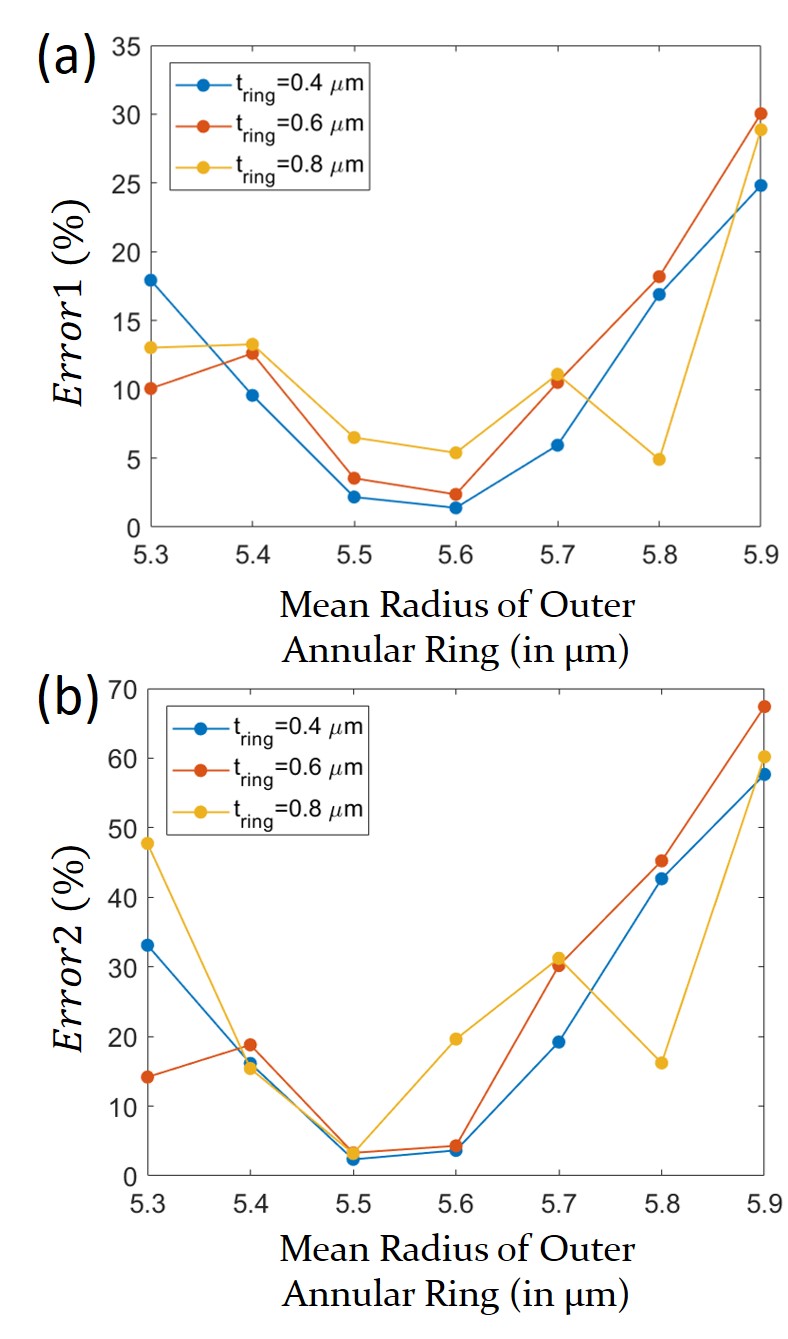}
    \caption{Minimum estimation errors (a) $(\frac{P_1-P_{1,est}}{P_1})_{min}\%$ and (b) $(\frac{P_2-P_{2,est}}{P_2})_{min}\%$ for $P_1$ and $P_2$ in the range 0 $kg m^{-2}$ to 10 $kg m^{-2}$ for different mean radii of outer ring $\frac{R_3+R_4}{2}$. Here, $\frac{R_1+R_2}{2}=2.9 \mu m$, $\frac{R_3+R_4}{2}=5.3-5.9 \mu m$, $R=8 \mu m$ and $t_{ring}=R_2-R_1=R_4-R_3=0.4,0.6,0.8 \mu m$.}
\end{figure}

Now, we can consider a test case wherein analyte A is positioned within the first annular ring, defined by an inner radius ($R_1$) of 2.9 $\mu m$ and an outer radius ($R_2$) of 3.3 $\mu m$, and analyte B is placed within the second annular ring, bounded by an inner radius ($R_3$) of 5.6 $\mu m$ and an outer radius ($R_4$) of 6.0 $\mu m$. The shifts in the resonance frequencies $(\frac{\Delta f_n}{f_n})$ of the graphene nanodrum are governed by equations (13) and (14). Here, $P_1$ and $P_2$ are the surface mass densities of analytes A and B respectively.

\begin{equation}
    \frac{\Delta f_1}{f_1} = lP_1 + mP_2
\end{equation}

\begin{equation}
    \frac{\Delta f_2}{f_2} = nP_1 + oP_2
\end{equation}

By inverting equations (13) and (14), we can estimate $P_1$ and $P_2$ as

\begin{equation}
\begin{bmatrix}
P_1 \\[6pt]
P_2
\end{bmatrix}
=
\begin{bmatrix}
l & m \\[6pt]
n & o
\end{bmatrix}
^{-1}
\begin{bmatrix}
\frac{\Delta f_1}{f_1} \\[6pt]
\frac{\Delta f_2}{f_2}
\end{bmatrix}
\end{equation}

Here, $l$, $m$, $n$, and $o$ are slopes of $(\frac{\Delta f_n}{f_n})$ vs $P_n$ obtained when either $P_1$ or $P_2$ is set to 0. These slopes are depicted in Fig. 6. Fig. 6(a) depicts the variation of $\frac{\Delta f_1}{f_1}$ having slope $l$ and $\frac{\Delta f_2}{f_2}$ having slope $n$ when $P_2 = 0$. Fig. 6(b) depicts the variation of $\frac{\Delta f_1}{f_1}$ having slope $m$ and $\frac{\Delta f_2}{f_2}$ having slope $o$ when $P_1 = 0$. Using data from Fig. 6, the values of $l$, $m$, $n$, and $o$ can be obtained, and henceforth can be used to estimate $P_1$ and $P_2$ based on equation (15). The values of $f_1$ and $f_2$ for different combinations of $P_1$ and $P_2$ are obtained using finite-element simulations conducted in COMSOL Multiphysics, and are depicted in Fig. 7. Since we have $
\begin{bmatrix}
l & m \\[4pt]
n & o
\end{bmatrix} = 
\begin{bmatrix}
-0.2363 & -1.461 \\[4pt]
-1.461 & -0.018
\end{bmatrix}
$, its inverse can be calculated as $\begin{bmatrix}
l & m \\[4pt]
n & o
\end{bmatrix}^{-1} = 
\begin{bmatrix}
0.00845 & -0.685 \\[4pt]
-0.685 & 0.1109
\end{bmatrix}$. \\

The estimated values and the corresponding percentage errors are presented in Figs. 8 and 9 respectively. The maximum observed error in estimating $P_1$ and $P_2$ are 1.8206 $\%$ and 2.0269 $\%$ respectively. This validates the robustness and reliability of the approach for this configuration. Also, we repeated this analysis for different geometrical configurations. Fig. 10 shows that the mass estimation error for $P_1$ is consistently lower than that for $P_2$ for multiple runs. Also, for any specific mean radius of inner ring $(\frac{R_1+R_2}{2})$ ranging from 2.8 $\mu m$ to 3.2 $\mu m$, the mass estimation errors for $P_1$ and $P_2$ are always lower for the mean radius of outer ring $(\frac{R_3+R_4}{2})$ in the range between 5.5 $\mu m$ and 5.6 $\mu m$. Here, we kept the thickness of the rings $(R_2-R_1)$ and $(R_4-R_3)$ constant at 0.4 $\mu m$. Fig. 11 now shows the mass estimation errors for varying thicknesses of the rings viz. 0.4 $\mu m$, 0.6 $\mu m$ and 0.8 $\mu m$. It can be seen that the estimation errors are lower for thinner rings.

The results thus indicate the importance of the node of one vibration mode to coincide with the antinode of another vibration mode, and vice versa as depicted in Fig. 5. This will ensure minimal modal interference
and assure that each analyte is associated with one dominant mode. When an analyte is placed near an antinode, it causes a larger shift in the resonance frequency, which makes it easier to detect and estimate the added mass. In contrast, if the analyte is placed near a node, the vibration in that area is minimal. Hence, even if mass is added, the frequency shift is small and harder to measure accurately, resulting in higher error. Since different mean radii correspond to different radial positions on the nanodrum, some radii naturally align better with antinodes in the vibration pattern. Hence, for these radii, an accurate detection of mass in the vibration modes is possible with lower estimation error. \\ 

Apart from the position of the annular rings, the thickness of each ring also plays an important role in determining the accuracy of mass estimation. Thicker rings cover a larger radial area on the nanodrum, which may not constitute pure antinodal/nodal regions. Hence, it can be harder to isolate the contribution of specific analytes leading to higher estimation errors. In contrast, thinner rings are more spatially localized and can be carefully placed within regions that correspond strictly to antinodes in one of the vibration modes. This focused placement ensures that the analyte induces more localized frequency shift, improving detection accuracy. However, if the ring is too thin, the overall influence on the frequency may become too weak to be detected reliably. Therefore, there exists a critical ring thickness that is large enough to cause a measurable frequency shift but small enough to avoid overlap with low-sensitive nodal regions. Our simulations indicate that a ring thickness of 0.4 $\mu m$ offers a good balance between sensitivity and selectivity for mass sensing, enabling accurate mass estimation with minimal cross-mode interference.

\section{Conclusions}

In this work, we propose and validate a computational framework for dual-analyte inertial imaging using a circular graphene nanodrum resonator with concentric annular mass distributions. By analytically modeling the vibrational mode shapes and coupling them with frequency shift data obtained from finite element simulations, we accurately estimate the individual mass densities of two spatially separated analytes. The least error in mass estimation is consistently obtained when the mean radius of the outer annular ring is set between 5.5 $\mu m$ and 5.6 $\mu m$. We also observe that thinner annular rings improve estimation accuracy by localizing the mass distribution and reducing cross-mode interference, highlighting the importance of spatial selectivity and ring geometry in the design. In the future, this approach can be extended to the simultaneous estimation of multiple masses $>2$ by tracking the resonance frequencies of multiple modes of vibration. This approach can also possibly be incorporated in nonlinear mechanical devices for exploring novel utilities \cite{ganesan2016observation}\cite{ganesan2017phononic}\cite{ganesan2018excitation}\cite{yang2021persistent}\cite{keskekler2022symmetry}\cite{kecskekler2023multimode}.


\printbibliography

\end{document}